# Performance Testing Effort at the ATM Forum: An Overview

Raj Jain and Gojko Babic, The Ohio State University

## Abstract

The Test working group at the ATM Forum is developing a specification for performance testing of ATM switches and networks. The emphasis is on the user perceived frame-level performance. This paper explains what is different about this new effort and gives its status.

## Introduction

ATM technology is now being deployed in operational networks. Most of the specifications required for operation have been developed. This includes signaling (UNI 4.0), routing (PNNI 1.0), traffic management (TM 4.0), numerous physical layer, network management, and testing specifications. As the technology moves from the laboratories to the field, the users have a need to benchmark and compare various ATM switches and other devices. The Devil's DP dictionary [5] defines performance benchmarking as follows:

**Benchmark** *v. trans.* To subject (a system) to a series of tests in order to obtain prearranged results not available on competitive systems.

This definition is not very far from truth. In the absence of a performance testing standard, each vendor is free to use whatever metric the vendor chooses and to measure those in an arbitrary manner. This can lead to confusion among buyers and users of the technology and can hurt the technology. It is, therefore, important to develop a set of standard performance metrics and precisely define their measurement and reporting procedure.

In October of 1995, we, therefore, made a proposal for starting the performance testing specification at the ATM Forum [3]. The forum members have been enthusiastic about this and considerable progress has been made since then. At every bimonthly meeting of the Forum, joint meetings of Traffic Management and Test working groups have been held to discuss Performance Testing. A baseline document is being prepared. Although much work remains to be done, goal of this paper is to provide an advance insight into this effort to ATM switching and monitoring product designers and other readers[1].

## Cell-level vs. Frame-level Metrics

One of the key distinguishing features of this new effort is its emphasis on frame-level metrics. In the past, the performance of ATM equipment as well as the quality of service was defined in terms of cell-level metrics. Cell loss ratio (CLR), cell delay variation (CDV), and cell transfer delay (CTD) are examples of cell-level metrics. Unfortunately, cell-level metrics do not reflect the performance as experienced (or desired) by the end users. Most users have frames to send and for the same cell loss ratio, the user perceived performance can be very different depending upon whether the cells dropped belong to a few frames or whether they belong to many frames. The user is more interested in frame loss ratio. Here, the term "frame" refers to AAL or higher layer protocol data unit (PDU).

A similar argument can be made against other cell-level metrics. For example, a video user sending 30 frames/second would like to receive complete frames every 33 milliseconds. It does not matter whether the cells belonging to a frame arrive together or arrive regularly spaced. Thus, it is the frame delay variation that matters and not the cell delay variation.

Frame-level metrics are also helpful in allowing ATM technology to be compared with non-ATM technology. For example, given a traffic pattern, a user could compare the performance of several network design alternatives - some of which may be ATM based and others may be non-ATM based.

Based on these arguments, the ATM Forum Test working group members decided to start the work on defining frame-level metrics.

## Goals of the ATM Forum Work

The objective of the ATM Forum work on performance testing is to enhance the marketability of ATM technology and equipment. The Forum will define metrics that will help compare various ATM equipment in terms of performance.

---

[1] The views expressed here are solely those of the authors and are not meant to reflect the official position of the ATM Forum.

The metrics should be independent of switch architecture. For example, "percentage of frames cut-through without delay" applies only to switches with cut-through feature and is not meaningfully applied to other (store and forward) switches. Such architecture-dependent metrics will not be defined.

The Forum plans to develop precise methodologies for measuring the metrics so that anyone can measure and produce the same result. The methodologies will include specific configurations, traffic patterns, and measurement procedures.

## Non-Goals of the ATM Forum Work

The ATM Forum does not intend to perform any measurements itself. Any vendor, user, or independent laboratory can use the methodologies and metrics developed by the Forum. The Forum does not intend to certify any particular measurements or laboratories.

Also, the Forum does not intend to set any thresholds of required performance. What frame loss ratio or frame delay variation is acceptable is left to the user and the supplier. Generally, there is a tradeoff between cost and acceptable performance. The users may accept equipment that is slow if it is cheap, while they may expect faster performance from expensive equipment. Different vendors will try to provide different cost-performance tradeoff points and such differentiation is generally good for a technology.

## Metrics

Most of the metrics discussed here apply to a single switch as well as a network of switches. Therefore, we use "system under test" or just "system" to refer to the device(s) being tested. A partial list of the metrics includes: throughput, frame latency, throughput fairness, frame loss ratio, maximum frame burst size, call establishment latency and application goodput. A brief overview of these metrics follows.

## Throughput

Three different frame-level throughput metrics are defined. **Lossless throughput** is the maximum rate at which none of the offered frames is dropped by the system. **Peak throughput** is the maximum rate at which the system operates regardless of frames dropped. The maximum rate can actually occur when the loss is not zero. **Full-load throughput** is the rate at which the system operates when the input links are loaded at 100% of their capacity.

A model graph of throughput vs. input rate is shown in Figure 1. Level $x$ defines the lossless throughput, level $y$ defines the peak throughput and level $z$ defines the full-load throughput.

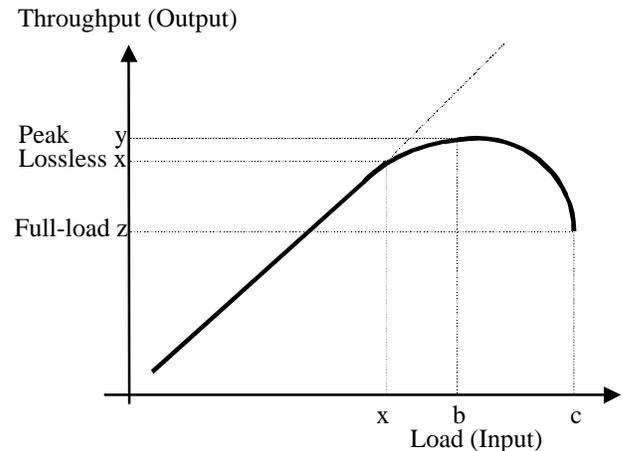

**Figure 1**: Peak, lossless and full-load throughput

The lossless throughput is the highest load at which the count of the output frames equals the count of the input frames. The peak throughput is the maximum throughput that can be achieved in spite of the losses. The full-load throughput is the throughput of the system at 100% load on input links. Note that the peak throughput may equal the lossless throughput in some cases. Only frames that are received completely without errors are included in frame-level throughput.

Throughput is expressed in effective bits/sec, counting only bits from AAL payloads excluding the overhead introduced by the ATM technology and transmission systems. This is preferred over specifying it in frames/sec or cells/sec. Frames/sec requires specifying the frame size. The throughput values in frames/sec at various frame sizes cannot be compared without first being converted into bits/sec. Cells/sec is not a good unit for frame-level performance since the cells are not seen by the user.

Before starting measurements, a number of VCCs (or VPCs), called foreground VCs, are established through the system. Foreground VCs are used to transfer only the traffic whose performance is being measured. That traffic is referred as the foreground traffic.

Foreground traffic is specified by the type of foreground VC, connection configuration, service class, arrival patterns, frame length, and input rate.

Foreground VCs can be permanent or switched, virtual path or virtual channel connections, established between ports on the same network module on the switch, or between ports on different network modules, or between ports on different switching fabrics.

A system with *n* ports is tested for the following connection configurations:

- **n-to-n straight**: Input from one port exits to another port. This represents almost no path interference among VCs. There are *n* VCs.

- **n-to-(n–1) full cross**: Input from each port is divided equally to exit on each of other (*n*–1) ports. This represents intense competition for the switching fabric by VCs. There are $n \times (n-1)$ VCs.

- **n-to-m partial cross**: Input from each port is divided equally to exit on other *m* ports ($1 \leq m \leq n-1$). This represents partial competition for the switching fabric by VCs. There are $n \times m$ VCs. Note that n-to-n straight and n-to-(n–1) full cross are special cases of n-to-m partial cross with *m*=1 and *m*=*n*–1, respectively.

- **k-to-1**: Input from *k* ($1 < k < n$) ports is destined to one output port. This stresses the output port logic. There are *k* VCs.

- **1-to-(n–1)**: Input from one port is multicast to all other output ports. This tests the multicast performance of the switch. There is only one VC.

Different connection configurations are illustrated in Figure 2, where each configuration includes one ATM switch with four ports, with their input components shown on the left and their output components shown on the right.

The following service classes, arrival patterns and frame lengths for foreground traffic are used for testing:

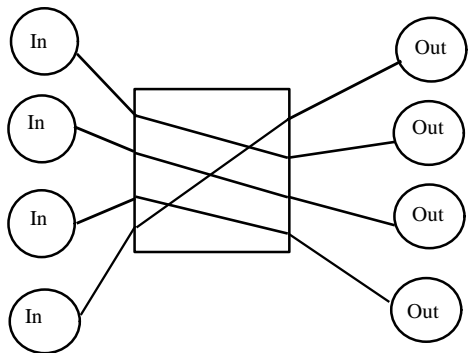
a. n-to-n straight: *n* VCs; *n*=4

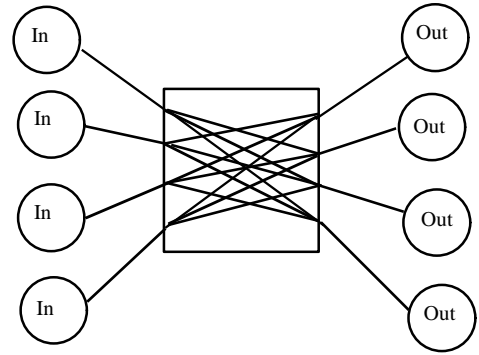
b. n-to-(n–1) full cross: $n \times (n-1)$ VCs; *n*=4

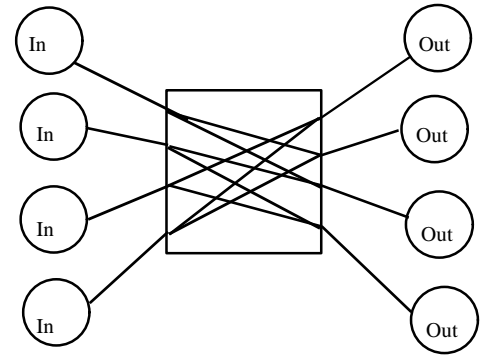
c. n-to-m partial cross: $n \times m$ VCs; *n*=4, *m*=2

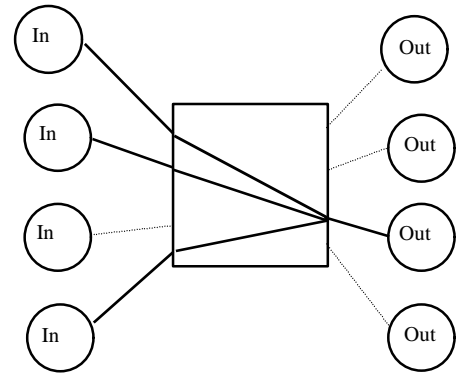
d. k-to-1: *k* VCs; *k*=3

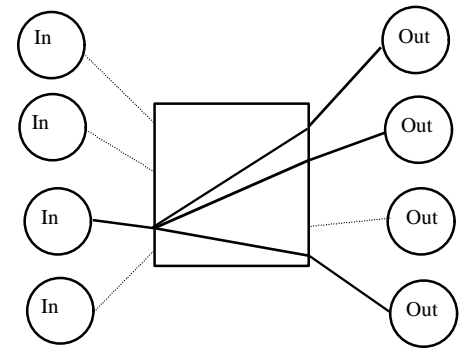
e. 1-to-(n–1): one multicast VCs

**Figure 2**: Connection configurations for throughput measurements

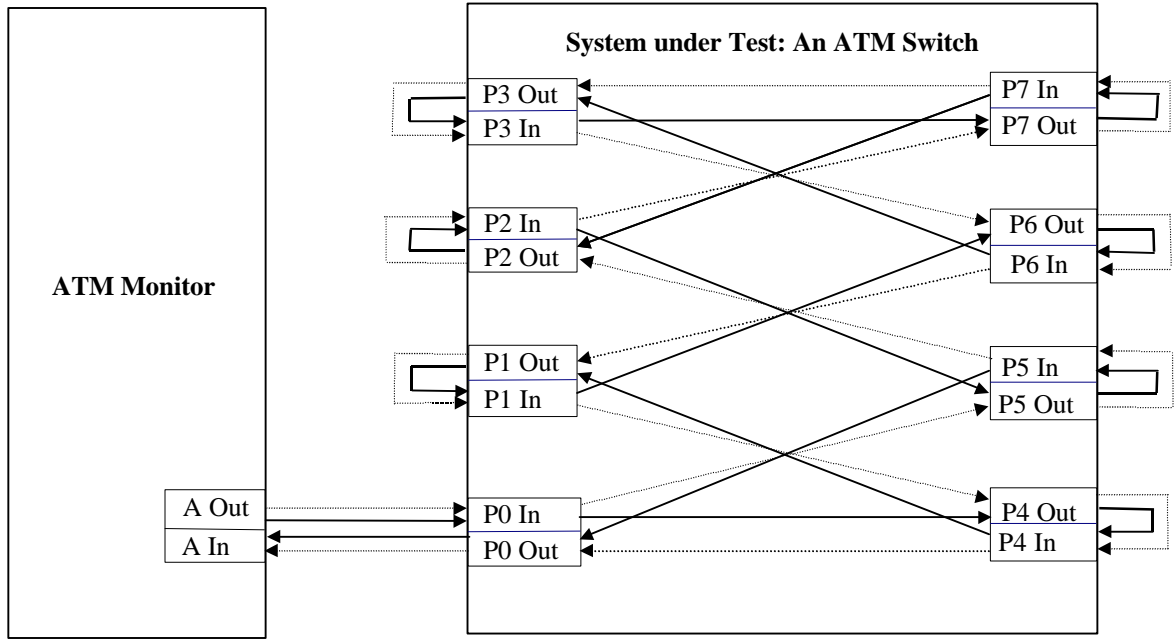

**Figure 3**: A scaleable test configuration for throughput measurements using only one generator/analyzer pair with 8-port switch and a 8-to-2 partial cross connection configuration

- UBR service class: Traffic consists of equally spaced frames of fixed length. Measurements are performed at AAL payload size of 64 B, 1518 B, 9188 B and 64 kB. Variable length frames and other arrival patterns (e.g. self-similar) are under study.

- ABR service class is under study.

Higher priority traffic like VBR or CBR can act as background traffic. Details of background traffic characteristics have not yet been defined.

The input rate of foreground traffic is expressed in the effective bits/sec, counting only bits from AAL payloads excluding the overhead introduced by the ATM technology and transmission systems.

It is obvious that testing larger systems, e.g., switches with larger number of ports, could require very extensive (and expensive) measurement equipment. Hence, we introduce scalable test configurations for throughput measurements that require only one ATM monitor with one generator/analyzer pair. Figure 3 presents a sample test configuration for an ATM switch with 8 ports in an 8-to-2 partial cross connection configuration. The configuration emulates 16 foreground VCs.

There is one link between the ATM monitor and the switch. The other seven ports have external loopbacks. A loopback on the given port causes the frames transmitted over the output of the port to be received by the input of the same port.

The test configuration in Figure 3 assumes two network modules in the switch, with switch ports P0-P3 in one network module and switch ports P4-P7 in the another network module. In this case, foreground VCs are always established from a port in one network module to a port in another network module.

This connection configuration could be more demanding on the system than the cases where each VC uses ports in the same network module. An even more demanding case could be when foreground VCs use different fabrics of a multi-fabric switch.

Similar approaches can be used for n-to-n straight, n-to-n full cross and other types of n-to-m partial cross connection configurations, as well as for larger switches.

## Frame Latency

MIMO latency (Message-In Message-Out) is a general definition of the latency that applies to an ATM switch or a group of ATM switches and it is defined as follows:

MIMO latency = min{LILO latency, FILO latency – NFOT}

where:
- LILO latency = Time between
    the last-bit entry and the last-bit exit
- FILO latency = Time between
    the first-bit entry and the last-bit exit

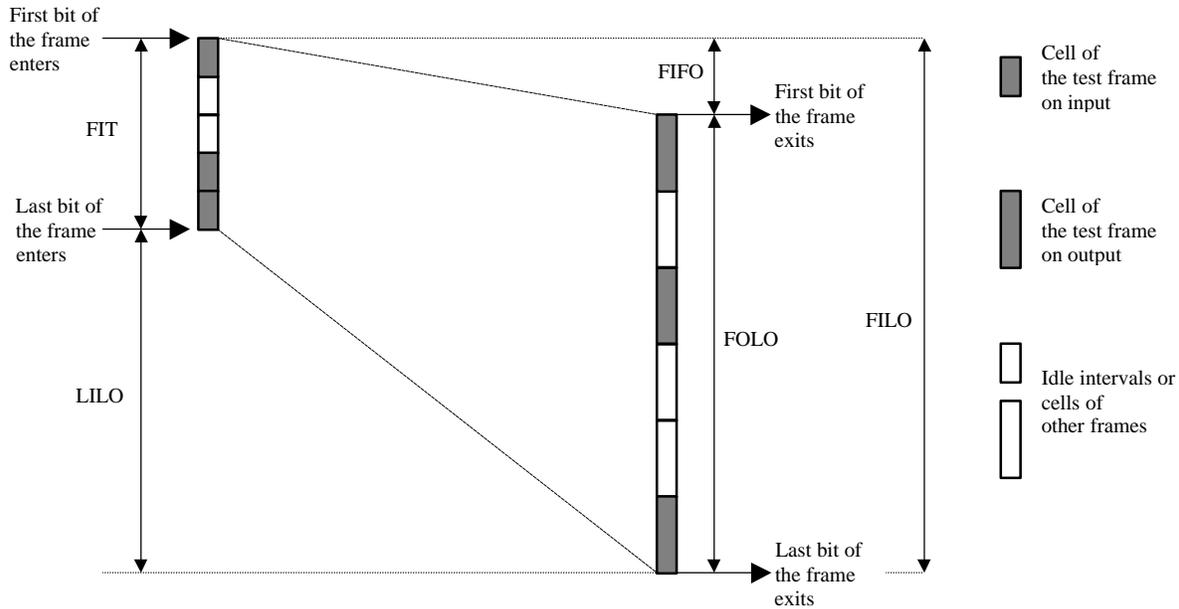

**Figure 4:** Latency metrics

- NFOT (Nominal Frame Output Time) =
  FIT × input rate / output rate
- FIT (Frame Input Time) = Time between
  the first-bit entry and the last-bit entry

An explanation of MIMO latency and its justification are presented in [4].

The MIMO is a general definition that applies even when the frames are discontinuous at the input and/or output or when the input and output rates are different.

To measure MIMO latency for a given frame, the times of the following three events should be recorded:

1. First-bit of the frame enters into the system
2. Last-bit of the frame enters into the system
3. Last-bit of the frame exits from the system

Time between events 1 and 3 is FILO latency and the time between events 2 and 3 is LILO latency. Also, NFOT can be calculated given the frame pattern on input (which includes a number of cells of the test frame and duration of idle periods, and/or number of cells from other frames, if any, between the first cell and the last during input transmission of the test frame) and (input and output) link rates. Then, substituting LILO latency, FILO latency and NFOT in the MIMO latency formula would give the frame latency of the system.

Contemporary ATM monitors provide measurement data only at the cell level, e.g. cell transfer delay (CTD), and cell inter-arrival time. This data is sufficient to calculate MIMO frame latency as follows.

If the input link rate is less than or equal to the output link rate, then:

MIMO latency = Last cell's transfer delay –
  (Last cell's input transmit time + Monitor overhead)

where:
- the cell input transmit time is the time to transmit one cell into the input link. It can be easily calculated.
- the monitor overhead is the overhead introduced by the ATM monitor when measuring CTD and it is usually non-zero. It can be calculated as the difference between the measured cell transfer delay for the case of a closed loop on the ATM monitor and the theoretical value for the cell transmit time plus any propagation delay.

Thus, to calculate MIMO latency when the input link rate is less than or equal to the output link rate, it is sufficient to measure the transfer delay of the last cell of a frame.

If the input link rate is greater than or equal to the output link rate, then:

MIMO latency = FIFO latency + FOLO time – NFOT

where:
- FIFO latency = First cell's transfer delay – (First cell's output transmit time + Monitor overhead)
- FOLO time = First cell to last cell inter-arrival time + Last cell's output transmit time
- the cell output transmit time is the time to transmit one cell into the output link. Again, it can be easily calculated.

Thus, to calculate MIMO latency when the input link rate is greater than or equal to the output link rate, it is necessary to measure the first cell transfer delay and the inter-arrival time between the first cell and the last cell of a frame.

For MIMO latency measurements, it is first necessary to establish one VCC or VPC used only by foreground traffic (the foreground VC), and a number of VCCs or VPCs used only by background traffic (background VCs). Then, the background traffic is generated. When flow of the background traffic has been established, the foreground traffic is generated. After the steady state flow of foreground traffic has been reached, required times and/or delays needed for MIMO latency calculation are recorded for $p$ consecutive frames from the foreground traffic, while the flow of background traffic continues uninterrupted. Here $p$ is a parameter.

Let $M_i$ be the MIMO latency of the ith frame. Note that MIMO latency is considered to be infinite for lost or corrupted frames. The mean and standard errors of the measurement are computed as follows:

Mean MIMO latency = $(\Sigma M_i) / p$

Standard deviation of MIMO latency =
$(\Sigma (M_i - \text{mean MIMO latency})^2) / (p-1)$

Standard error =
standard deviation of MIMO latency $/ p^{1/2}$

Given the mean and standard errors, the users can compute a $100(1-a)$-percent confidence interval as follows:

$100(1-a)$-percent confidence interval =
(mean $- z \times$ standard error, mean $+ z \times$ standard error)

Here, $z$ is the $(1-a/2)$-quantile of the unit normal variate. For commonly used confidence levels, the quantile values are as follows:

| Confidence | a | Quantile |
|---|---|---|
| 90% | 0.1 | 1.615 |
| 99% | 0.01 | 2.346 |
| 99.9% | 0.001 | 3.291 |

MIMO latency depends upon several characteristics of the foreground traffic. These include the type of foreground VC, service class, arrival patterns, frame length, and input rate.

The foreground VC can be permanent or switched, virtual path or virtual channel connection, established between ports on the same network module, or between ports on different network modules, or between ports on different fabrics.

For the UBR service class, the foreground traffic consists of equally spaced frames of fixed length. Measurements are performed at AAL payload sizes of 64 B, 1518 B, 9188 B and 64 kB. Variable length frames and other arrival patterns (e.g. self-similar) are under study. ABR service class is also under study.

The input rate of foreground traffic is expressed in effective bits/sec, counting only bits from AAL payloads excluding the overhead introduced by the ATM technology and transmission systems.

The first measurement run is performed at the lowest possible foreground input rate (for the given test equipment). For later runs, the foreground load is increased up to the point where losses occur or up to the full foreground load (FFL). FFL is equal to the lesser of input or output link rate used by the foreground VC.

Background traffic characteristics that affect frame latency are the type of background VC, connection configuration, service class, arrival patterns (if applicable), frame length (if applicable), and input rate.

Like the foreground VC, background VCs can be permanent or switched, virtual path or virtual channel connections, established between ports on the same network module, or between ports on different network modules, or between ports on different fabrics. To avoid interference on the traffic generator/analyzer equipment, background VCs are established in such a way that they do not use the input link or the output link of the foreground VC in the same direction.

For a system with $w$ ports, the background traffic can use $(w-2)$ ports, not used by the foreground traffic, for both input and output. The input port of foreground traffic can be used as an output port for background traffic. Similarly, the output port of foreground traffic can be used as an input port for background traffic. Overall, background traffic can use an equivalent of $n=w-1$ ports. The maximum background load (MBL) is defined as the sum of rates of all links, except the one used as the input link for the foreground traffic.

A system with $w$ $(=n+1)$ ports is measured for the following background traffic connection configurations:

- n-to-n straight, with $n$ VCs,
- n-to-(n–1) full cross, with $n \times (n-1)$ VCs,
- n-to-m partial cross, $1 \leq m \leq n-1$, with $n \times m$ VCs,
- 1-to-(n–1), with one multicast VC.

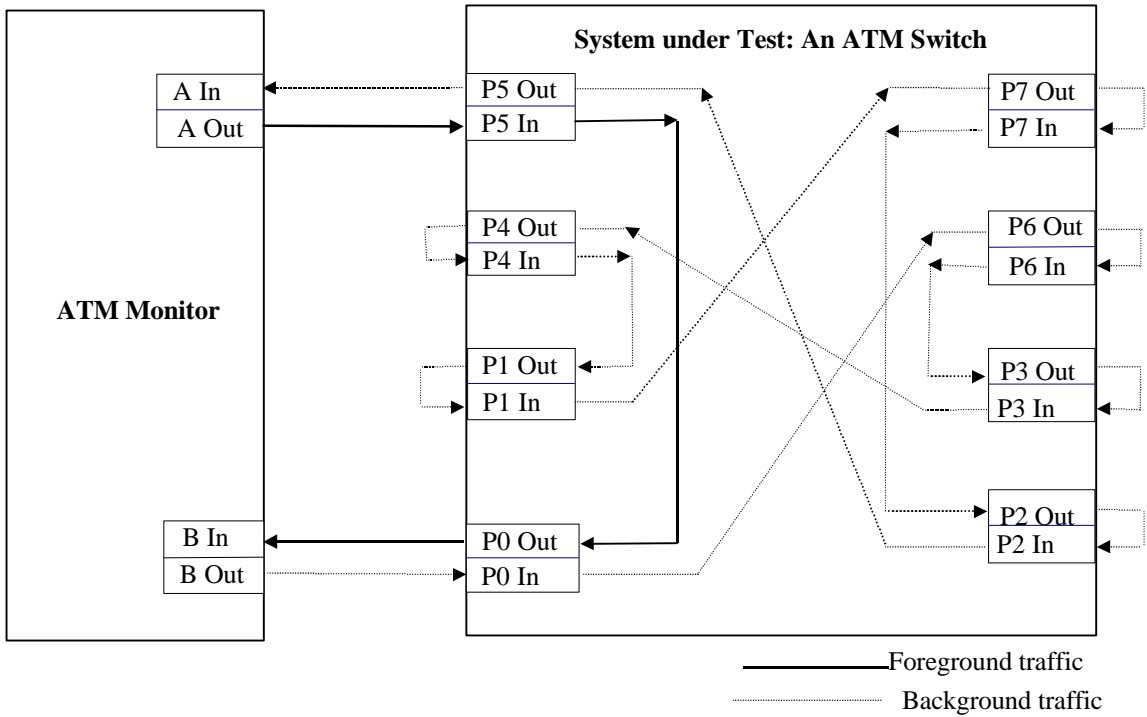

**Figure 5**: A scalable test configuration for measurement of MIMO latency using only two generator/analyzer pairs with 8-port switch and a 7-to-7 straight connection configuration for the background traffic

These configurations are the same as those shown earlier in Figure 2.

The following service classes, arrival patterns (if applicable) and frame lengths (if applicable) are used for the background traffic:

- UBR service class: Traffic consists of equally spaced frames of fixed length. Measurements are performed at AAL payload size of 64 B, 1518 B, 9188 B and 64 kB. This is a case of bursty background traffic of priority equal to or lower than that of the foreground traffic. Variable length frames and other arrival patterns (e.g. self-similar) are under study.

- CBR service class: Traffic consists of a contiguous stream of cells at a given rate. This is a case of non-bursty background traffic of priority higher than that of the foreground traffic.

- VBR and ABR service classes are under study.

Scaleable test configurations for MIMO latency measurements require only one ATM monitor with two generator/analyzer pairs. Figure 5 presents the test configuration with an ATM switch with eight ports ($w$=8). There are two links between the ATM monitor and the switch, and they are used in one direction by the background traffic and in the another direction by the foreground traffic, as indicated. The other six ($w$–2) ports of the switch are used only by the background traffic and they have external loopbacks.

Figure 5 shows a 7-to-7 straight connection configuration for the background traffic. The n-to-(n–1) full cross and n-to-m partial cross connection configurations can also be similarly implemented.

The test configuration shown assumes two network modules in the switch with switch ports P0-P3 in one network module and switch ports P4-P7 in the another network module. Here, the foreground VC and background VCs are established between the two ports in different network modules.

It should be noted that in these test configurations, if all link rates are not identical, it is not possible to generate background traffic (without losses) equal to MBL. The maximum background traffic input rate in such cases equals ($n$–1)×lowest link rate. Only if all link rates are identical, it is possible to obtain MBL level without losses in the background traffic.

## Throughput Fairness

Given $n$ contenders for the resources, throughput fairness indicates how far the actual individual allocations are from the ideal allocations. In the most general case, the ideal allocation is defined by the max-min allocation[2] to various contending virtual circuits. For the simplest case of *n* VCs sharing a link with a total throughput *T*, the throughput of each VC should be *T/n*.

If the actual measured throughputs of *n* VCs sharing a system (a single switch or a network of switches) are found to be $\{T_1, T_2, ..., T_n\}$, where the optimal max-min throughputs should be $\{\hat{T}_1, \hat{T}_2, ..., \hat{T}_n\}$, then the fairness of the system under test is quantified by the "fairness index" computed as follows [1]:

$$\text{Fairness index} = (\Sigma x_i)^2 / (n \times \Sigma x_i^2)$$

where, $x_i = T_i/\hat{T}_i$ is the relative allocation to *i*th VC.

This fairness index has the following desirable properties:

- It is dimensionless. The units used to measure the throughput (bits/sec, cells/sec, or frames/sec) do not affect its value.
- It is a normalized measure that ranges between zero and one. The maximum fairness is 100% and the minimum 0%. This makes it intuitive to interpret and present.
- If all $x_i$'s are equal, the allocation is fair and the fairness index is one.
- If *n–k* of *n* $x_i$'s are zero, while the remaining *k* $x_i$'s are equal and non-zero, the fairness index is *k/n*. Thus, a system which allocates all its capacity to 80% of VCs has a fairness index of 0.8 and so on.

Throughput fairness is quantified by the fairness index for each of the throughput experiments in which there are either multiple VCs or multiple input/output ports. Thus, it applies to all three throughput measures (lossless, peak, and full-load), all connection configurations and all traffic patterns. No additional experiments are required for throughput fairness. The detailed results obtained for the throughput tests are analyzed to compute the fairness.

The throughput tests are run several times for a specified duration. The fairness is computed for each individual run. Let $F_i$ be the fairness index for the *i*th run, then the mean fairness is computed as follows:

$$\text{Mean fairness} = (\Sigma F_i) / \text{Number of repetitions}$$

Note that fairness index is not limited to throughput. It can be applied to other metrics, such as latency. However, extreme unfairness in latency is expected to appear as unfairness in throughput and vice versa.

---

[2] Other policies can be used but must be specified.

## Frame Loss Ratio

Frame loss ratio is defined as the fraction of frames that are not forwarded by a system due to lack of resources. Partially delivered frames are considered lost.

Frame loss ratio = (input frame count –
              output frame count) / input frame count

There are two frame loss ratio metrics that are of interest to a user:

- **Peak throughput frame loss ratio**: It is the frame loss ratio at the input load corresponding to the peak throughput.
- **Full-load throughput frame loss ratio**: It is the frame loss ratio at the input load corresponding to the full-load throughput.

These metrics are related to the throughput:

Frame loss ratio =
              (input rate – throughput) / input rate

Thus, no additional experiments are required for frame loss ratios. These can be derived from tests performed for throughput measurements provided the input rates are recorded.

The throughput experiments are repeated a specified number of times. If $FLR_i$ is the frame loss ratio for the *i*th run:

$FLR_i = (\text{input rate}_i – \text{throughput}_i) / \text{input rate}_i$

Since frame loss ratio is a "ratio," its average cannot be computed via straight summation [2]. The average frame loss ratio (FLR) for multiple runs is computed as follows:

$FLR = (\Sigma \text{input rate}_i – \Sigma \text{throughput}_i) / \Sigma \text{input rate}_i$

## Maximum Frame Burst Size

Maximum frame burst size (MFBS) is the maximum number of frames that a source end system can send at the peak rate through a system without incurring any loss.

MFBS measures the data buffering capability of the system and its ability to handle back-to-back frames.

Many applications and transport layer protocol drivers often present a burst of frames to AAL for transmission. For such applications, maximum frame burst size provides a useful indication.

This metric is particularly relevant to the UBR service category since UBR sources are always allowed to send a burst at the peak rate. ABR sources may be throttled down to a lower rate if the switch runs out of buffers.

MFBS is expressed in octets of AAL payload. This is preferred over number of frames or cells because the former requires specifying the frame size and the latter is not very meaningful for a frame-level metric. Also, the number of cells has to be converted to octets for use by AAL users.

It may be useful to indicate the frame size for which MFBS has been measured. If MFBS is found to be highly variable with frame size, a number of common AAL payload field sizes such as 64 B, 536 B, 1518 B, and 9188 B may be used.

The number of frames sent in the burst is increased successively until a loss is observed. The maximum number of frames that can be sent without loss are reported as MFBS.

## Call Establishment Latency

For short duration VCs, call establishment latency is an important part of the user perceived performance. Informally, the time between submission of a call setup request to a network and the receipt of the connect message from the network is defined as the call establishment latency. The time lost at the destination while the destination was deciding whether to accept the call is not under network control and is, therefore, not included in call setup latency (See Figure 6).

Thus, the sum of the latency experienced by the setup message and the resulting connect message is the call setup latency.

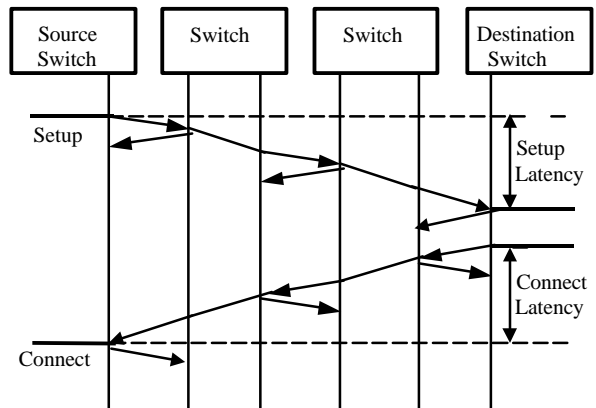

**Figure 6**: Call establishment latency

The main problem in measuring these latencies is that both these messages span multiple cells with intervening idle cells. Unlike X.25, frame relay, and ISDN networks, the messages in ATM networks are not contiguous. Therefore, the MIMO latency metric is used. Thus,

Call establishment latency =
   MIMO latency for setup message +
   MIMO latency for connect message

The call establishment latency as defined above applies to any network of switches. In practice, it has been found that the latency depends upon the number of switches and the number of PNNI group hierarchies traversed by the call. It is expected that measurements will be conducted on multiple switches connected in a variety of ways. In all cases, the number of switches and number of PNNI group hierarchies traversed are indicated.

## Application Goodput

Application goodput captures the notion of what an application sees as useful data transmission in the long term. Application goodput is the ratio of packets (frames) received to packets (frames) transmitted over a measurement interval.

The application goodput (AG) is defined as:

AG = frames received in the measurement interval /
   frames transmitted in the measurement interval

Traditionally, goodput is measured in bits per sec. However, we are interested in a non-dimensional metric and are primarily interested in characterizing the useful work derived from the expended effort rather than the actual rate of transmission. While the application-goodput is intended to be used in a single-hop mode, it does have meaningful end-to-end semantics over multiple hops.

*Notes:*

- This metric is useful when measured at the peak load. The number of transmitted frames is varied over a useful range from 2000 frames per second (fps) through 10000 fps at a nominal frame size of 64 B. Frame sizes are also varied through 64 B, 1518 B, and 9188 B to represent small, medium, and large frames respectively. Note that the frame sizes specified do not account for the overhead of accommodating the desired frame transmission rates over the ATM medium.

- The measurement interval should be chosen to be large enough to accommodate the transmission of the largest packet (frame) over the connection and small enough to track short-term variations of the average goodput.

- It is important to not include network management frames and/or keep-alive frames in the count of received frames.
- There should be no changes of handling buffers during the measurement interval.
- The results are to be reported as a table for the three different frame sizes.

## The OSU ATM Benchmarking Laboratory

Any vendor or user can run the benchmarks and tests developed by the ATM Forum. Still there is a need for an independent measuring organization that can conduct the tests and publish results on a regular basis. The Ohio State University ATM benchmarking laboratory will play this role. The role of this laboratory for ATM testing will be similar to that at Harvard for router and LAN switch testing. We have been awarded funding by the National Science Foundation and the State of Ohio for this laboratory.

## Summary

The ATM forum Test and Traffic Management groups are jointly working on defining a set of standard performance metrics and tests. The key distinguishing feature of this work is that it considers the user perceived performance and therefore uses frame-level metrics rather then the cell-level metrics of the past.

In this paper, we provided a brief overview of several metrics that are being defined. The metrics, their definitions and tests are currently being refined.

## References[3]

---

[3] All our past ATM forum contributions and presentations are available on-line at http://www.cis.ohio-state.edu/~jain/

## Additional Reading

**Raj Jain** is a professor of Computer and Information Sciences at the Ohio State University in Columbus, Ohio. Prior to joining the university in April 1994, he was a Senior Consulting Engineer at Digital Equipment Corporation, Littleton, Massachusetts, where he was involved in design and analysis of many computer systems and networks including VAX Clusters, Ethernet, DECnet, OSI, FDDI, and ATM networks. Currently he is very active in the Traffic Management working group of ATM Forum and has influenced its direction considerably. He is also the editor of the ATM Forum's Performance Testing Specification.

He received B.E. in Electrical Engineering from A.P.S. University, Rewa, India 1972, M.E. in Automation from Indian Institute of Science, Bangalore, 1974, and Ph.D. in Applied Math (Computer Science) from Harvard University, Cambridge, Mass, 1978. He is the author of "Art of Computer Systems Performance Analysis," Wiley 1991 and "FDDI Handbook: High-Speed Networking with Fiber and Other Media," Addison Wesley, 1994. He is an IEEE Fellow and an ACM Fellow. For further information and publications, please see http://www.cis.ohio-state.edu/~jain/

**Gojko Babic** received his graduate degree in electrical engineering from University of Sarajevo in 1972, his M.S. degree in computer science from Florida Institute of Technology in 1975, and his Ph.D. in computer science from the Ohio State University in 1978.

In the period 1983-1990, he was a leader and principal investigator of the program Energonet, the largest and the most important research and development project in computer networking in the former Yugoslavia. Main results were the public X.25 network for Bosnia-Herzegovina and X.25 network in China for the Ministry of Public Security. In the period 1988-1992, he was the director of Computer Networking Department at IRIS-Computer, a division of Energoinvest Corporation and an associate professor at the University of Sarajevo. He is currently a senior lecturer at the Ohio State University.

He is an IEEE senior member and an ACM member since 1975.